\documentstyle[11pt,newpasp,epsf]{article}

\begin{document}

\title{New insights in high-resolution spectroscopy: a wide theoretical
library of R=500\,000 stellar spectra}

\author{E. Bertone}
\affil{Instituto Nacional de Astrof\'\i sica, Optica y Electr\'onica, A.P. 51 y 216, 72000 Puebla, Mexico}

\author{A. Buzzoni}
\affil{Telescopio Nazionale Galileo, A.P. 565, E-38700 Santa Cruz de La Palma, Canary Islands, Spain; and Osservatorio Astronomico di Brera, Milano, Italy}

\author{L.H. Rodr\'\i guez-Merino and M. Ch\'avez}
\affil{Instituto Nacional de Astrof\'\i sica, Optica y Electr\'onica, A.P. 51 y 216, 72000 Puebla, Mexico}

\begin{abstract}

We present an extended theoretical library of over 800 synthetic stellar
spectra, covering energy distribution in the optical range ($\lambda =
3500-7000$~\AA), at inverse resolution R=500\,000. The library, based on the
ATLAS\,9 model atmospheres, has been computed with the SYNTHE code developed
by R.\ L.\ Kurucz. The grid spans a large volume in the fundamental parameters
space (i.e.\ $T_{\rm eff}$, $\log{g}$, [M/H]), and can be profitably
applied to different research fields dealing both with the study of single
stars and stellar aggregates, through population synthesis models. A
complementary project, in progress, will extend the wavelength range to the
ultraviolet, down to 850~\AA, at an inverse resolution of R=50\,000.

\end{abstract}



\section{Introduction}

New-generation spectrographs, at the major ground-based telescopes, have begun 
to pour in the hard-disks of astronomers' computers an increasing mass of
high-quality spectroscopic data. Inverse resolutions as high as $\lambda /
\Delta\lambda = 100\,000$ can now be easily attained, at least for the
brightest ($V \la 15$ mag) objects in the sky, and this is pushing the
observation of local and extragalactic stellar systems to an ever unimagined
resolution level. The outstanding performance of instruments like UVES and
VIRMOS at the ESO Very Large Telescope, or HIRES and SARG at the Keck
Observatory and Telescopio Nazionale Galileo, respectively, urges therefore
theoretical tools of comparable accuracy level in order to consistently match
and analyse such a huge amount of observational data.

In this framework, and to help filling the gap, we undertook a long-term
project aimed at providing the community with a systematic theoretical library
of high-resolution stellar spectra (virtually the largest sample currently
available in the literature) in the optical range ($\lambda = 3500 \to
7000$~\AA) and at an inverse resolution of $R=500\,000$.

\section{The optical grid of synthetic spectra}

This first set of spectral energy distributions relied on the SYNTHE code of
Kurucz (1993), using ATLAS\,9 model atmospheres (Kurucz 1995) as
input. Computations have been carried out at the Brera Observatory in Milan
(Italy). 
\index{Observatories!Brera}
The input models follow the classical
approximations of steady-state, homogeneous, LTE,  plane-parallel layers, with
a microturbulence velocity of 2 km/s and a mixing-length value
$\ell/H_p=1.25$. They make use of an improved treatment of convective
overshooting in the calculation of the transfer equations (Castelli, Gratton,
\& Kurucz 1997).

Over 46 million absorption lines are accounted for in the code, including all
atomic elements at different ionization states and the most important diatomic
molecules. All line data were extracted from the Kurucz' (1992)
original database, with the major improvement of the TiO contribution, for
which we adopted the new list of lines computed by Schwenke (1998).

The wavelength interval spans the whole optical range, from the Balmer break
to H$\alpha$. All the passbands of the Lick/IDS spectrophotometric indices
(Worthey et al. 1994) are included, as well as the
high-resolution index set defined by Rose (1994). Spectra are sampled
at variable wavelength step maintaining a constant resolution
($\Delta \lambda/\lambda = 2 \times 10^{-6}$).

The library extends to effective temperatures as hot as $T_{\rm eff} =
50\,000$~K, and provides a suitable match to O~$\to$~K spectral types. M
stars, cooler than $T_{\rm eff} < 4000$~K, are however still lacking in our
dataset given the missing contribution of tri-atomic molecular opacity in the
input model atmospheres (Kurucz 1992).

\begin{figure}[!h]
\begin{center}
\plotone{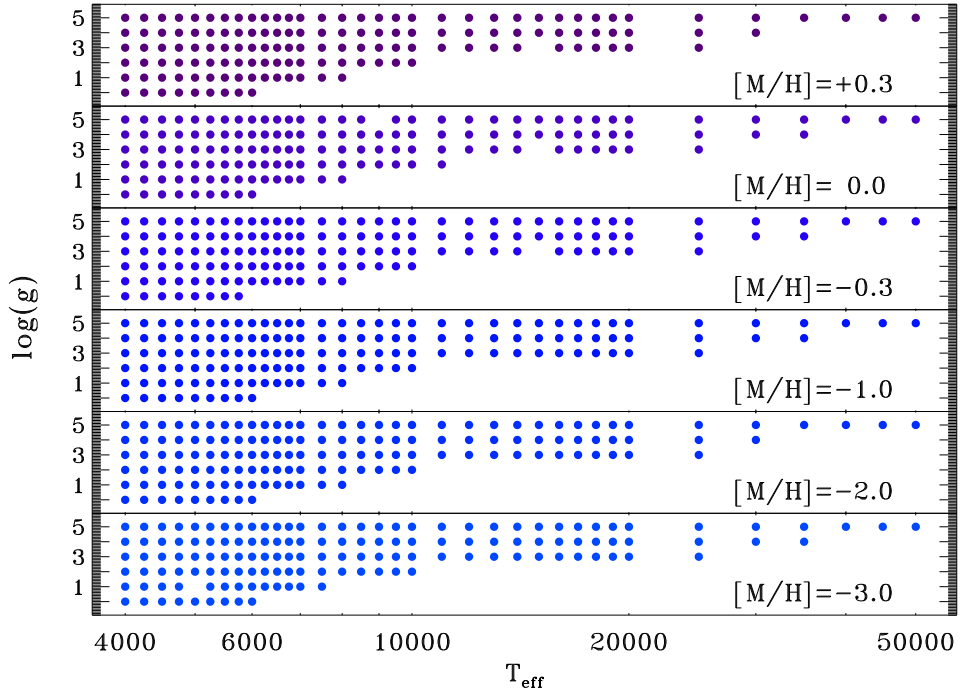}
\caption{The optical library of synthetic high-resolution spectra. Each panel
shows the model grid for fixed metallicity. The whole set consists of 832
SEDs.}
\label{fig:bertonee1}
\end{center}
\end{figure}

\begin{figure}[!t]
\begin{center}
\plotone{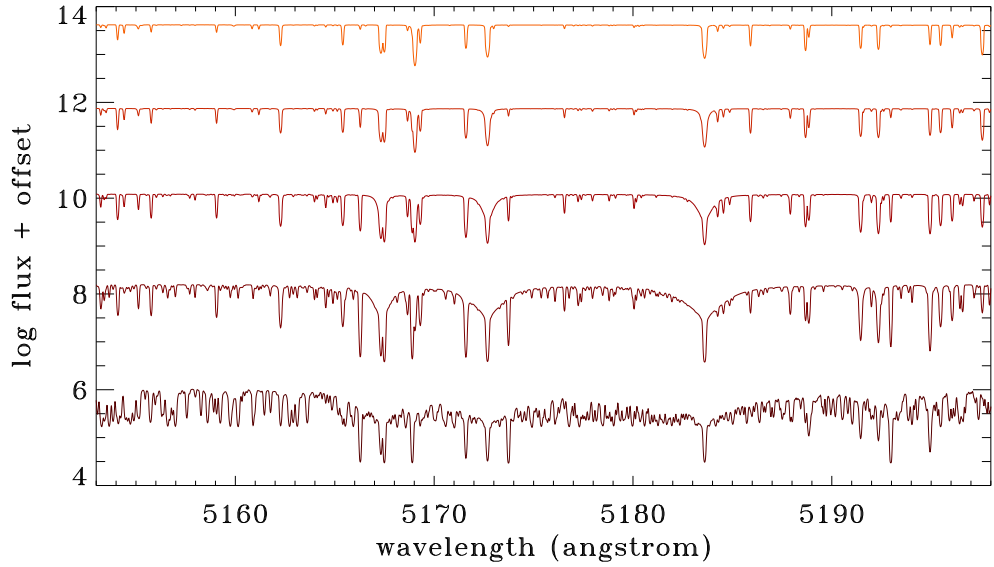}
\caption{An illustrative sample of the synthetic spectra with solar
metallicity, spanning the wavelength region around the atomic Mg\,$b$
triplet. Increasing temperatures (botton to top) from $T_{\rm eff} = 4000 \to
8000$~K at step of 1000~K are explored, with fixed gravity $\log{g} = 4$.}
\label{fig:bertonee2}
\end{center}
\end{figure}

\begin{figure}[!h]
\begin{center}
\plotone{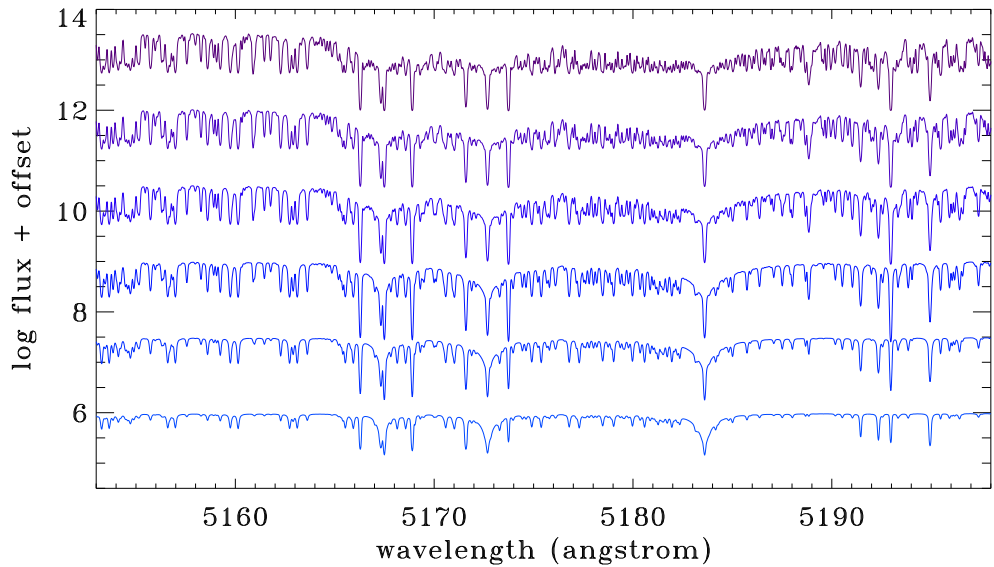}
\caption{Same as in Fig.~\ref{fig:bertonee2}, but with varying metallicity,
for $T_{\rm eff} = 4000$~K and $\log{g} = 4$. Metallicity values are ${\rm
[M/H]} = -3.0, -2.0, -1.0, -0.3,\, 0.0, +0.3$ (bottom to top, respectively).}
\label{fig:bertonee3}
\end{center}
\end{figure}

Gravity spans a wide range across the H-R diagram ($5 \geq \log{g} \geq 0$),
namely from dwarf (MK class V) to supergiant stars (MK class I), while
metallicity accounts for stars at the two extreme edges of the Galaxy [Fe/H]
distribution with $-3.0 < {\rm [M/H]} < +0.3$ as boundary limits. A total of
832 synthetic spectral energy distributions (SEDs) have been computed, as
summarized in Fig.~\ref{fig:bertonee1} and Table~\ref{tab:grid}.
An illustrative example of the model output, sampling the spectral emission
around the atomic Mg\,$b$ triplet, is shown in Fig.~\ref{fig:bertonee2} and
\ref{fig:bertonee3}, with varying physical parameters.

\begin{figure} 
\begin{center}
\plotone{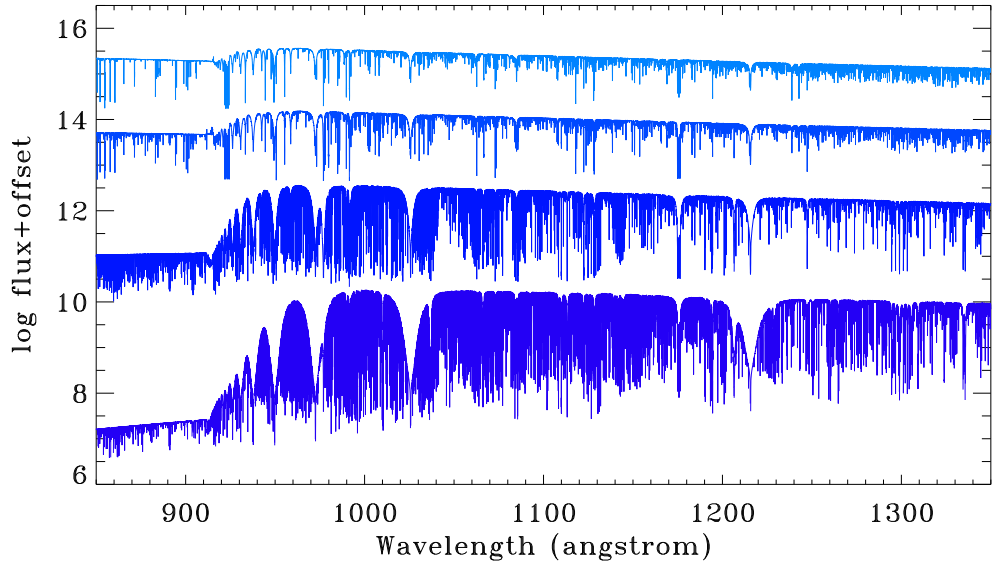}
\caption{A sequence of high temperature spectra longward of the Lyman limit
from the UV library. The models have fixed gravity ($\log{g} = 5$) and solar
chemical composition, while the effective temperature increases (bottom to
top) from 20\,000 to 50\,000~K at step of 10\,000~K.}
\label{fig:bertonee4}
\end{center}
\end{figure}

A complementary project, which will extend our analysis to the ultraviolet
interval, between 850--4750~\AA\ with an inverse resolution $R = 50\,000$, is
in progress at INAOE and will be completed soon (Rodr\'\i guez-Merino
2002). The extended dataset consists of over 1000 SEDs  covering the
effective temperature range between 3500 and 50\,000~K at surface gravity
$\log{g}$ = 1.0~$\to$~5.0 and chemical compositions ${\rm [M/H]} = +0.5,\, 0.0,
-0.5$ and $-1.5$ (Rodr\'\i guez-Merino et al. 2001). In
Fig.~\ref{fig:bertonee4} we display a subset of spectra computed for solar
chemical composition, $\log{g}=5$ and different effective temperatures.

\begin{table}[!t]
\caption{The properties of the optical library of synthetic spectra.}
\label{tab:grid}
\begin{center}
\begin{tabular}{ll}
\tableline \noalign{\medskip}
Code:             & SYNTHE (Kurucz 1993) \\
Input models:     & ATLAS9 (Kurucz 1995) \\
Wavelength range: & $3500 \; \leq \; \lambda \; \leq \; 7000 \;$~\AA  \\
Resolution:       & $ R = \lambda / \Delta \lambda = 500\,000$ \\
Wavelength step:  & $ 0.007\; \leq \; \Delta\lambda \; \leq \; 0.014 \;$~\AA \\
Wavelength points: & 346\,645 \\
Total absorption lines: & $\sim 46$ millions \\
Molecules:   & C$_2$, CN, CO, CH, NH, OH, MgH, SiH, \\
             & H$_2$, SiO, TiO \\
Total number of spectra: & 832 \\
Effective temperature:   & $4000 \; \leq \; T_{\rm eff} \; \leq \; 50000 \;$~K  \\
Surface gravity:         & $0.0 \; \leq \; \log{g} \; \leq \; 5.0$ \\
Metallicity:             & $-3.0 \; \leq \; {\rm [M/H]} \; \leq \; +0.3$ \\ 
\noalign{\medskip}
\tableline \tableline
\end{tabular}
\end{center}
\end{table}


\begin{figure}
\begin{center}
\plotone{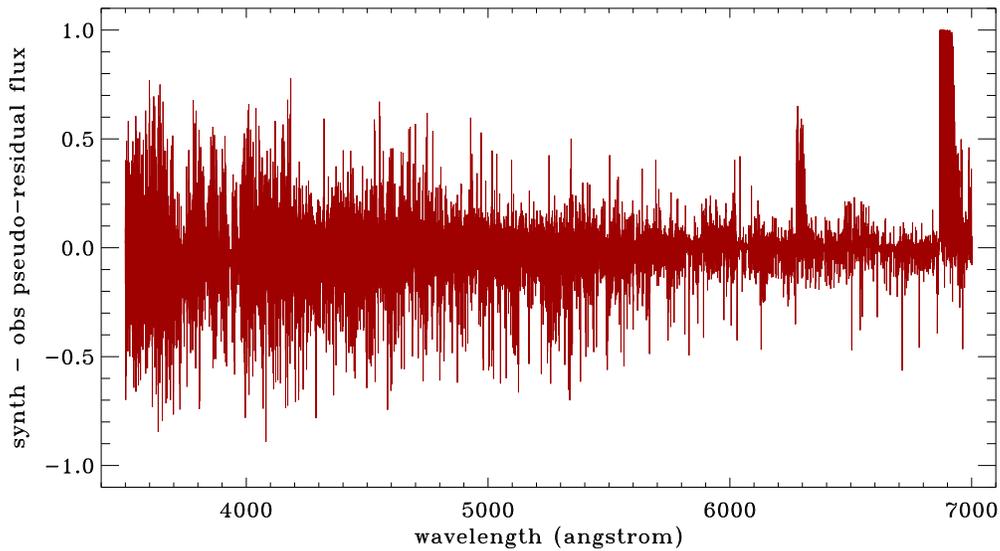}
\caption{Difference in the pseudo-residual flux between the synthetic and the
observed spectrum of the Sun (Kurucz et al. 1984) in the spectral
region 3500--7000~\AA. The features about 6300 and 6900~\AA\ are due to
telluric absorption bands in the observed spectrum.}
\label{fig:bertonee5}
\end{center}
\end{figure}

\section{The solar spectrum validation}

A comparison with the solar spectrum represents a first mandatory test in
order to assess the degree of reliability of our theoretical library. We
present in Fig.~\ref{fig:bertonee5} the match of the Solar Flux
Atlas by Kurucz et al. (1984) and the corresponding synthetic solar
spectrum, which we computed at similar resolution (i.e.\ $R=522\,000$).
The input model atmosphere is from the ATLAS\,9 and assumes $T_{\rm eff} =
5777$~K, $\log{g} = 4.4377$, a microturbulent velocity of 1.0 km/s, and
chemical abundance according to Anders \& Grevesse (1989).

The figure is made up by a dense sequence of positive and negative spikes,
which mainly track the residual flux differences in the absorption line
profiles. While positive spikes mainly originate from observed lines that are
not present in the synthetic model, the negative peaks come from theoretical
absorption lines that are too weak or undetected in the observed
spectrum. Note, as a stricking feature in the plot, that both positive and
negative spikes tend to decrease at longer wavelength, and no systematic drift
is present in the data distribution.
Fig.~\ref{fig:bertonee5} gives also a direct measure of the
intrinsic uncertainty in the input physics of the models; excluding the
telluric bands, clearly affecting the residual distribution about 6300 and
6900~\AA, the rms value in the plot is $\pm 0.086$.


\begin{references}
\reference Anders, E., \& Grevesse, N. 1989, Geochim. Cosmochim. Acta, 53, 197
\reference Castelli, F., Gratton, R. G., \& Kurucz, R. L. 1997, \aap, 318, 841
\reference Kurucz, R. L. 1992, in IAU Symp. 149, The Stellar Populations of
Galaxies, ed. B. Barbuy \& A. Renzini, (Dordrecht: Kluwer), 225
\reference Kurucz, R. L. 1993, CD-ROM No. 18, SYNTHE spectrum synthesis
programs and line data
\reference Kurucz, R. L. 1995, CD-ROM No. 13, ATLAS9 Stellar Atmosphere
Programs and 2 km/s Grid, revised
\reference Kurucz, R. L., Furenlid, I., Brault, L., \& Testerman, L. 1984,
National Solar Observatory Atlas, Sunspot, New Mexico: National Solar
Observatory
\reference Rodr\'\i guez-Merino, L.H. 2002, Ph. D. Thesis, in preparation
\reference Rodr\'\i guez-Merino, L. H., Ch\'avez, M., Buzzoni, A., Bertone, E.
2001, in proc. of New Quests in Stellar Astrophysics: A Link between Stars
and Cosmology, ed. M. Chavez, A. Bressan, A. Buzzoni \& D. Mayya, (Dordrecht:
Kluwer), 39
\reference Rose, J. A. 1994, \aj, 107, 206
\reference Schwenke, D. W. 1998, Faraday Discuss., 109, 321
\reference Worthey, G., Faber, S. M., Gonz\'ales, J. J., \& Burnstein,
D. 1994, \apjs, 94, 687
\end{references}
\end{document}